\documentclass[sigconf,nonacm]{acmart}

\setcopyright{none}
\renewcommand\footnotetextcopyrightpermission[1]{}
\usepackage{amsmath}
\usepackage{booktabs}
\usepackage{array}
\usepackage{enumitem}
\usepackage{tikz}
\usepackage{quantikz}

\begin{document}

\title{Integration Amplitude Loading as a General Primitive for Quantum
Monte Carlo Integration in Finance}

\author{Tamuz Danzig}
\email{tamuz@classiq.io}
\affiliation{%
  \institution{Classiq Technologies}
  \city{Tel Aviv}
  \country{Israel}
}

\begin{abstract}
Monte~Carlo integration is central to quantitative finance. Pricing derivatives
and measuring risk both reduce to averaging a payoff or loss over many scenarios,
a computation that is demanding at the accuracy the industry requires. Quantum
amplitude estimation (QAE), together with its iterative variant (IQAE), can reach
a target accuracy with quadratically fewer evaluations than classical Monte~Carlo. In
practice, though, the cost of these algorithms is governed by \emph{amplitude
loading}, the step that writes the integrand into a qubit amplitude, and most
existing loading methods produce deep circuits that consume a large amount of
resources. We argue that \emph{integration amplitude loading} (IAL), which was
developed for pricing rainbow and autocallable options, is really a general tool
for evaluating expectations rather than a construction tied to a particular
payoff. Its central element is a quantum comparator that carries out the
integration directly, so that loading the integrand reduces to a single
comparator acting on a cheaply prepared auxiliary register. For autocallable
options this lowers the T-depth of the payoff by about 50x relative to other
loading methods, and for other financial problems the method is likely to deliver
a comparable improvement over existing approaches. We describe the method without
reference to any specific payoff and show that it applies across quantitative
finance, with option pricing, value-at-risk and expected shortfall, and credit
risk among the most immediate applications. Risk measures are the most natural of
these, since a tail-risk figure is itself a tail integral of the loss
distribution.
\end{abstract}


\maketitle

\section{Introduction}
Monte~Carlo integration is the workhorse of quantitative finance. Pricing a
derivative, measuring market or credit risk, computing counterparty exposure, or
estimating a sensitivity all come down to the same operation: averaging a payoff
or a loss function over a distribution of scenarios. Reaching the accuracy that
practitioners demand can require a very large number of scenarios, which makes
this averaging one of the more expensive computations in the field. Quantum
amplitude estimation (QAE), and especially its iterative variant (IQAE), targets
precisely this cost. It attains a given accuracy using a number of evaluations
that scales with the inverse of the error, rather than the inverse square required
by classical Monte~Carlo, a quadratic
improvement~\cite{Montanaro2015,Brassard2002,Grinko2021}. The same routine
already underlies quantum algorithms for derivative
pricing~\cite{Rebentrost2018,Stamatopoulos2020}, risk
analysis~\cite{Woerner2019}, and credit risk~\cite{Egger2021}.

These algorithms share a common structure. First the probability distribution is
prepared as a quantum state. Then the integrand is loaded into the amplitude of a
target qubit. Finally, amplitude estimation reads off the expectation. It is the
middle step that is expensive. On fault-tolerant hardware the dominant cost is the
T-depth of the amplitude-loading circuit~\cite{Chakrabarti2021}, since T gates
require costly magic-state distillation and their sequential depth sets the
running time. Loading a non-trivial function with existing general-purpose methods
means either assembling it from quantum arithmetic and piecewise-polynomial
approximations or synthesizing it with quantum signal processing (QSP), and both
routes lead to a large T-depth.

Integration amplitude loading (IAL) was introduced to avoid this cost when
pricing rainbow options~\cite{Cibrario2024} and, in a refined form,
path-dependent autocallable options~\cite{Cibrario2025}, where it cut the T-depth
of the payoff by about 50x relative to other loading methods. That figure is
specific to the autocallable case, but the same mechanism is expected to give
significant improvements for other financial problems as well. Our claim in this
paper is that the technique is not really about options. What makes it work has
nothing to do with any particular payoff: a comparator carries out the
integration, so any expectation whose integrand is monotone, or can be broken into
monotone pieces, and whose increments can be prepared cheaply, is a candidate for
the same treatment.

The reason a single primitive reaches so much of finance is that its core
operation, integrating a function against a loaded distribution with one
comparator, is exactly what these problems require. Option pricing adds one more
common structure, the exponential map that takes log-returns to prices, which IAL
also handles cheaply. In concrete terms the method applies to exotic and
path-dependent options such as Asian, barrier, lookback, basket, and autocallable
contracts, as well as to linear products such as contracts for difference. It
applies to value-at-risk, conditional value-at-risk, and expected
shortfall~\cite{Woerner2019}, to credit-portfolio loss and economic
capital~\cite{Egger2021}, to counterparty exposure measures such as potential
future exposure~\cite{Jha2026}, to the computation of
Greeks~\cite{Stamatopoulos2022}, and to insurance and catastrophe tail
risk~\cite{Kirke2026}. Risk measures deserve particular mention, since a tail-risk
quantity is itself a tail integral of the loss distribution, which is exactly what
the comparator produces, and this is where we expect IAL to matter most.

The remainder of the paper restates IAL as a payoff-independent primitive
(Section~\ref{sec:primitive}), discusses the major financial use cases and the
quantities they estimate (Section~\ref{sec:reach}), and closes with a broader
outlook (Section~\ref{sec:discussion}).

\section{Background: QAE-based MC integration}
The quantity of interest is an expectation of an integrand $f$ over a
probability distribution $P$,
\begin{equation}
\mathbb{E}[f(X)]=\int f(x)\,dP(x),
\end{equation}
which QAE estimates with $\mathcal{O}(1/\varepsilon)$ oracle queries for target
error $\varepsilon$, against the classical Monte~Carlo cost of
$\mathcal{O}(1/\varepsilon^{2})$~\cite{Montanaro2015,Brassard2002}.
Let $X$ be discretized over $2^{n}$ points with probabilities $p(\omega)$ loaded
by a state-preparation operator~\cite{GroverRudolph2002}, and let
$\tilde f:\{0,\dots,2^n-1\}\to[0,1]$ be the integrand rescaled into $[0,1]$.
Amplitude loading builds an operator $\mathcal{A}$ acting as
\begin{align}
\mathcal{A}\!\sum_\omega\!\sqrt{p(\omega)}|\omega\rangle|0\rangle
=\;&\sum_\omega\sqrt{(1-\tilde f(\omega))\,p(\omega)}\,|\omega\rangle|0\rangle
\nonumber\\
+\;&\sum_\omega\sqrt{\tilde f(\omega)\,p(\omega)}\,|\omega\rangle|1\rangle ,
\end{align}
so that the probability of measuring the target qubit in $|1\rangle$ is
$a=\sum_\omega p(\omega)\tilde f(\omega)=\mathbb{E}[\tilde f]$. QAE estimates $a$
with $\mathcal{O}(1/\varepsilon)$ oracle queries~\cite{Brassard2002}. The original
algorithm builds this estimate through quantum phase estimation (QPE), which needs
an additional register and a long sequence of controlled amplifications. Iterative
QAE (IQAE) and related phase-estimation-free variants drop the QPE step entirely
and reach the same quadratic speedup with much shallower
circuits~\cite{Grinko2021,Suzuki2020}. Because QAE calls $\mathcal{A}$
$\mathcal{O}(1/\varepsilon)$ times, the cost of implementing $\mathcal{A}$ is what
decides whether a practical quantum advantage can be
reached~\cite{Chakrabarti2021,Herbert2022}.

\section{The Integration Amplitude Loading Primitive}
\label{sec:primitive}
\begin{figure}[t]
\centering
\begin{quantikz}[column sep=0.35cm,row sep=0.35cm]
\lstick{$|0\rangle^{\otimes n}_{r}$} & \gate{\text{Prep }\sqrt{g}} & \gate[3]{r\le x} & \qw \\
\lstick{$|x\rangle_{n}$}            & \qw                        &                  & \qw \\
\lstick{$|0\rangle$}               & \qw                        &                  & \meter{}
\end{quantikz}
\caption{The IAL primitive. The increment state $\sqrt{g}$ ($g=\Delta\tilde f$)
is prepared on an auxiliary register, off the critical path, and a single
comparator $r\le x$ integrates it, writing $F(x)=\sum_{r\le x}g(r)=\tilde f(x)$
into the target amplitude.}
\label{fig:ial}
\end{figure}
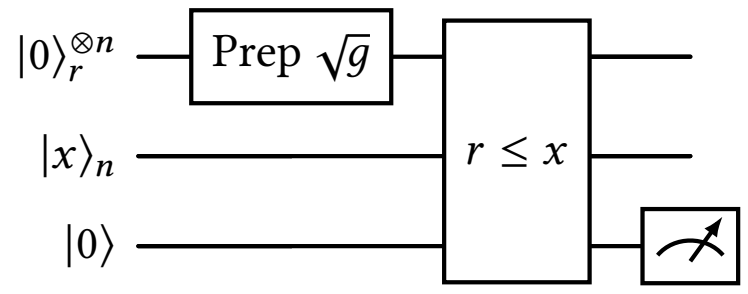

The method rests on a single observation, that a comparator can be used to carry
out an integration. Concretely, we introduce an $n$-qubit auxiliary register
prepared in the \emph{increment state} $\sum_r\sqrt{g(r)}\,|r\rangle$ together with
one target qubit, and apply a comparator that tests $r\le x$ against the value
register $|x\rangle$ (Fig.~\ref{fig:ial}). This produces
\begin{equation}
\sqrt{\textstyle\sum_{r=0}^{x}g(r)}\,|x\rangle|\psi\rangle|1\rangle
+\sqrt{\textstyle\sum_{r=x+1}^{2^n-1}g(r)}\,|x\rangle|\psi_\perp\rangle|0\rangle ,
\label{eq:integrate}
\end{equation}
i.e.\ the comparator accumulates the \emph{partial sum} $F(x)=\sum_{r\le x}g(r)$
into the target amplitude~\cite{Cibrario2024,Cibrario2025}.

Thus, to load a
monotone nondecreasing integrand $\tilde f$, one sets $g$ to its discrete
increment $g(r)=\tilde f(r)-\tilde f(r-1)\ge 0$. Then
$\sum_{r\le x}g(r)=\tilde f(x)$ exactly. \emph{Amplitude loading of $f$ becomes
amplitude preparation of $\Delta f$ followed by one comparator.}

What makes this construction inexpensive is that the increment-state preparation
acts only on the auxiliary register and can run \emph{in parallel} with
distribution loading and any arithmetic, leaving only the comparator on the serial
path. The cost of amplitude loading is therefore just that of a single controlled
comparator, which is logarithmic in the register size~\cite{Cibrario2025}.

The case that matters most in practice is the exponential increment
$g(r)=e^{ar}/Z$, with $Z=\sum_j e^{aj}$, which arises whenever asset prices are
log-normal. Its increment state is a \emph{product state}, so it is prepared by
$n$ \emph{parallel} single-qubit $R_y$ rotations with angles
\begin{equation}
\theta_i = 2\arctan\!\big(e^{a2^{i}/2}\big),
\label{eq:angles}
\end{equation}
and the comparator output of Eq.~\eqref{eq:integrate} then takes the closed form
\begin{equation}
\sqrt{\frac{e^{a(x+1)}-1}{e^{a2^{n}}-1}} .
\label{eq:expamp}
\end{equation}

The practical consequence is a much cheaper payoff circuit. For the autocallable
benchmark analyzed in~\cite{Cibrario2025}, IAL reduces the payoff-component T-depth
by about 50x relative to other loading methods, with further reductions in T-count
and qubit count.

Two mild requirements are enough for the method to apply: (a) the integrand is
monotone on the integration variable, or can be split into monotone segments whose
comparator integrals are summed, and (b) the increment state $\sqrt{g}$ is
efficiently preparable, which is trivial
for exponential, affine, and other product-structured increments and otherwise
needs a single round of exact amplitude amplification~\cite{Cibrario2025}.
Non-monotone integrands are handled by piecewise decomposition, and multivariate
integrands by choosing a monotone reduction, such as $\arg\max$ in log-return
space, or by loading one dimension at a time.

\section{Sample of Financial Use Cases}
\label{sec:reach}
The primitive of Section~\ref{sec:primitive} applies to any financial quantity
that can be written as an expectation of a monotone, or piecewise-monotone,
function over a loaded distribution. We describe a representative sample of use
cases below, together with the quantity that IAL estimates in each.

The most familiar example is option pricing, where the price is the discounted
expectation of the payoff under the risk-neutral distribution. For a European call
this expectation is
$\mathbb{E}[\max(S_T-K,0)]$, and for a contract for difference it is the
expectation of the linear payoff $S_T-S_0$. Prices follow from log-returns through
an exponential map, so the payoff is naturally expressed in log-return space,
where IAL loads it with a single comparator and, for the exponential part, a set
of parallel $R_y$ rotations. The same construction extends to path-dependent
contracts: Asian options on the average price, barrier and knock-out options
defined by threshold crossings, lookback options on the running extremum, basket
and rainbow options on several assets, and autocallable structures built from a
sequence of conditional payoffs. Barriers and knock-out conditions are especially
convenient, because the comparator that performs the integration is the same
operation that tests the threshold.

A second broad area is risk measurement, where value-at-risk, conditional
value-at-risk, and expected shortfall summarize the tail of a portfolio loss
distribution~\cite{Woerner2019}. Each is a tail integral
of that distribution, namely the probability of a loss beyond a threshold or the
average loss in the tail. These quantities map onto IAL with little modification,
because the comparator evaluates exactly the cumulative loss probability that
defines them, and the loss transform is absorbed into the prepared increment
state. Tail risk is also where classical Monte~Carlo struggles most, since the
events that drive the answer are rare, so this is the setting where the method
should help the most.

Credit and counterparty risk fit the same mould. In credit-portfolio models the
quantity of interest is the distribution of losses from correlated defaults, from
which one reads the expected loss and the economic capital, the difference between
a high quantile of the loss and its mean~\cite{Egger2021}. Counterparty exposure
measures such as potential future exposure ask instead for a high quantile of the
future exposure of a derivative or portfolio~\cite{Jha2026}. Both are tail quantities of a loss or exposure
distribution and follow the same integration pattern as market risk.

Sensitivity analysis and insurance round out the sample. Computing the Greeks, the
sensitivities of a price to model parameters, usually relies on finite differences
or gradients that re-price the instrument many times~\cite{Stamatopoulos2022}, and
since each re-pricing reuses the same inexpensive loader, the savings from IAL
accumulate across the whole calculation. In insurance, catastrophe and
excess-of-loss covers require the expected loss above an attachment point, which is
a tail integral of the severity distribution~\cite{Kirke2026} and again matches
the primitive directly.

\section{Discussion and Outlook}
\label{sec:discussion}
As a general primitive, IAL reduces the design of a quantum Monte~Carlo circuit to
two conditions on the integrand: that it be monotone, at least in pieces, and that
its increments be preparable efficiently. When both conditions hold, amplitude
loading, which has been the bottleneck all along, becomes a parallel state
preparation followed by a single comparator of logarithmic depth.

Because so many financial calculations are expectations of monotone payoffs or
losses, a single efficient loader has broad reach. The same construction that
prices an option also evaluates a value-at-risk, an economic-capital figure, a
counterparty exposure, or an insurance tail loss, with only the loaded
distribution and the choice of comparator changing from one problem to the next.
This uniformity is valuable in practice, since a firm that implements IAL once can
reuse it across trading desks and risk functions, rather than building a custom
payoff circuit for each product. The benefit also compounds: amplitude estimation
evaluates the loader many times, and sensitivity analysis repeats the whole pricing
many more times, so an inexpensive loader turns into a large saving at the level of
a full workload. These gains matter most for the tail-risk problems that dominate
regulatory and economic-capital reporting, where classical Monte~Carlo is most
strained by rare events, and this is where we expect the method to have the
greatest impact.

\bibliographystyle{ACM-Reference-Format}
\bibliography{refs}

\end{document}